\documentclass[letterpaper]{article} 
\usepackage{aaai2026}  
\usepackage{times}  
\usepackage{helvet}  
\usepackage{courier}  
\usepackage[hyphens]{url}  
\usepackage{graphicx} 
\usepackage{natbib}  
\usepackage{caption} 
\usepackage{algorithm}
\usepackage{algorithmic}

\usepackage{times}
\usepackage{helvet}
\usepackage{courier}
\usepackage{subcaption}
\usepackage{booktabs}
\usepackage{multirow}
\usepackage{graphicx}
\usepackage{amssymb}
\usepackage{amsmath}
\usepackage{newfloat}
\usepackage{listings}
\DeclareCaptionStyle{ruled}{labelfont=normalfont,labelsep=colon,strut=off} 
\floatstyle{ruled}
\newfloat{listing}{tb}{lst}{}
\floatname{listing}{Listing}
\title{ParaMETA: Towards Learning Disentangled Paralinguistic Speaking Styles Representations from Speech}
\author {
    Haowei Lou\textsuperscript{\rm 1},
    Hye-young Paik\textsuperscript{\rm 1},
    Wen Hu\textsuperscript{\rm 1}, 
    Lina Yao\textsuperscript{\rm 1, \rm 2},
}
\affiliations {
    \textsuperscript{\rm 1}University of New South Wales \\
    \textsuperscript{\rm 2}CSIRO's Data61\\
    haowei.lou@unsw.edu.au, h.paik@unsw.edu.au, wen.hu@unsw.edu.au,lina.yao@unsw.edu.au
}

\begin{document}

\maketitle

\begin{abstract}
Learning representative embeddings for different types of speaking styles, such as emotion, age, and gender, is critical for both recognition tasks (e.g., cognitive computing and human-computer interaction) and generative tasks (e.g., style-controllable speech generation). In this work, we introduce \textbf{ParaMETA}, a unified and flexible framework for learning and controlling speaking styles directly from speech. Unlike existing methods that rely on single-task models or cross-modal alignment, ParaMETA learns disentangled, task-specific embeddings by projecting speech into dedicated subspaces for each type of style. This design reduces inter-task interference, mitigates negative transfer, and allows a single model to handle multiple paralinguistic tasks such as emotion, gender, age, and language classification.
Beyond recognition, ParaMETA enables fine-grained style control in Text-To-Speech~(TTS) generative models. It supports both speech- and text-based prompting and allows users to modify one speaking styles while preserving others. Extensive experiments demonstrate that ParaMETA outperforms strong baselines in classification accuracy and generates more natural and expressive speech, while maintaining a lightweight and efficient model suitable for real-world applications.
\end{abstract}

\begin{links}
    \link{Code}{https://github.com/haoweilou/ParaMETA}
\end{links}

\section{Introduction}
Understanding and modelling speaking styles from speech is important for many real-world applications. In applications like affective computing and human-computer interaction, being able to recognize a speaker’s emotion, age, or gender helps systems respond in a more personalized and appropriate way~\cite{schuller2011recognising,li2013automatic,mairesse2007personage}. In speech generation applications, the ability to control speaking styles allows for the creation of more varied and expressive speech, which is especially useful in applications such as filmmaking, animation, virtual assistants, and conversational robots~\cite{skerry2018towards,wang2018style,ao2021speecht5}.

Learning effective representations of speaking styles is a critical foundation for both classification and generation tasks. In classification settings, the objective is to maximize the discriminability of embeddings across different speaking styles, such as distinguishing between happy and sad speech, so that the classifier can accurately identify each class. Cross-entropy loss remains a standard and effective optimization criterion for this purpose~\cite{mao2023cross}. However, speaking styles recognition encompasses multiple distinct tasks, such as emotion and age recognition, which involve different labeling and decision boundaries. Some studies address this by training separate models for each task~\cite{tripathi2020review,ravishankar2020prediction}, but this approach is often computationally expensive and difficult to scale. Alternatively,  multi-task models aim to address this by sharing parameters across tasks~\cite{qawaqneh2017age}. While more efficient, these models frequently suffer from inter-task interference, where learning one type of style negatively impacts the performance on others.
In contrast, generative applications not only require accurate modelling of speaking styles but also demand fine-grained controllability to allow flexible manipulation of these styles during synthesis. A variety of approaches have been proposed to control speaking styles in speech generation. Methods such as CosyVoice~\cite{du2024cosyvoice} and PromptTTS~\cite{guo2023prompttts} rely on descriptive text prompts to guide the speaking styles of generated speech. Other systems, including Spark-TTS~\cite{wang2025spark}, F5-TTS~\cite{chen2024f5}, and VALL-E~\cite{wang2023neural}, use speech prompts to extract style embeddings directly from reference speech.  More recently, UniStyle~\cite{zhu2024unistyle} has attempted to unify text- and speech-based style control by learning a joint latent embedding using Q-Former~\cite{li2023blip}. However, this tightly coupled design limits the controllability of speaking style. Specifically, Unistyle has been reported to generate speech that retains the characteristics of the reference speech, even when the text prompt specifies a conflicting style. 

CLAP (Contrastive Language-Audio Pretraining)\cite{elizalde2023clap} has emerged as a dominant approach for learning speech representations. It aligns speech and text into a unified embedding space, enabling both classification and generation tasks. Some audio and speech generation frameworks, such as AudioLDM~\cite{liu2023audioldm} and VoiceLDM~\cite{lee2024voiceldm}, leverage pretrained CLAP embeddings to guide the audio generation process. However, CLAP-based approaches face notable limitations in terms of speaking styles entanglement. 
Specifically, CLAP aligns speech with text by creating a unified embedding that contains all speaking styles described in the text captions, such as emotion, age, and gender.
This entanglement of speaking styles leads to inter-task interference and negative transfer, where dominant speaking styles overshadow others~\cite{zhang2022survey}. As a result, it becomes difficult to isolate and control individual style attributes, limiting flexibility in downstream applications. Furthermore, learning such a unified embedding space often requires large-scale models with high computational and memory demands~\cite{elizalde2023clap}.

To address the challenges of entangled style representations, inter-task interference, and limited controllability. We propose a \textbf{Para}lingustic \textbf{META} representation learning framework, \textbf{ParaMETA}, designed for modelling the speaking styles from speech.  Drawing inspiration from supervised contrastive learning~\cite{khosla2020supervised} and recent advances in vision-language modelling~\cite{zhang2025llava}, ParaMETA employs a two-stage embedding learning strategy. First, in the META embedding space, we group together speech samples with shared labels to maximize the utility of style labels in the dataset. 
Then, in the task-specific subspaces, we project the META embeddings into separate spaces for each task (e.g., emotion, gender, age) and optimize them independently. This ensures that embeddings belonging to the same class within a given task are closely clustered, regardless of other attributes. For example, in the emotion-specific space, speeches labeled as happy are placed near each other, even if they differ in gender or age.
This disentanglement effectively mitigates interference between tasks. For text-speech alignment, instead of using traditional joint embedding strategies, ParaMETA adopts an efficient projection-based approach inspired by LLaVA~\cite{zhang2025llava}. Specifically, text embeddings are projected directly into the speech embedding space through prototype-based alignment, enabling precise correspondence between modalities without incurring high computational costs. The main contribution of this work are: 
\begin{itemize}
    \item We propose ParaMETA, a new representation learning framework that effectively learns speaking styles from speech in a disentangled and structured manner.
    \item We propose to project and learn different types of speaking styles within their own task-specific spaces, effectively reducing inter-task interference and mitigating the problem of negative transfer.
    \item  We design a new approach for controlling speaking styles in TTS, which unifies both speech- and text-based prompting and enables fine-grained style control across different types of speaking styles.
    \item  We conduct comprehensive experiments showing that ParaMETA consistently outperforms baselines in classification accuracy and delivers superior control over speaking styles in speech generation tasks.
\end{itemize}

\section{Method}
\begin{figure*}[t!]
    \centering
    \begin{subfigure}[b]{0.36\linewidth}
        \includegraphics[width=\linewidth]{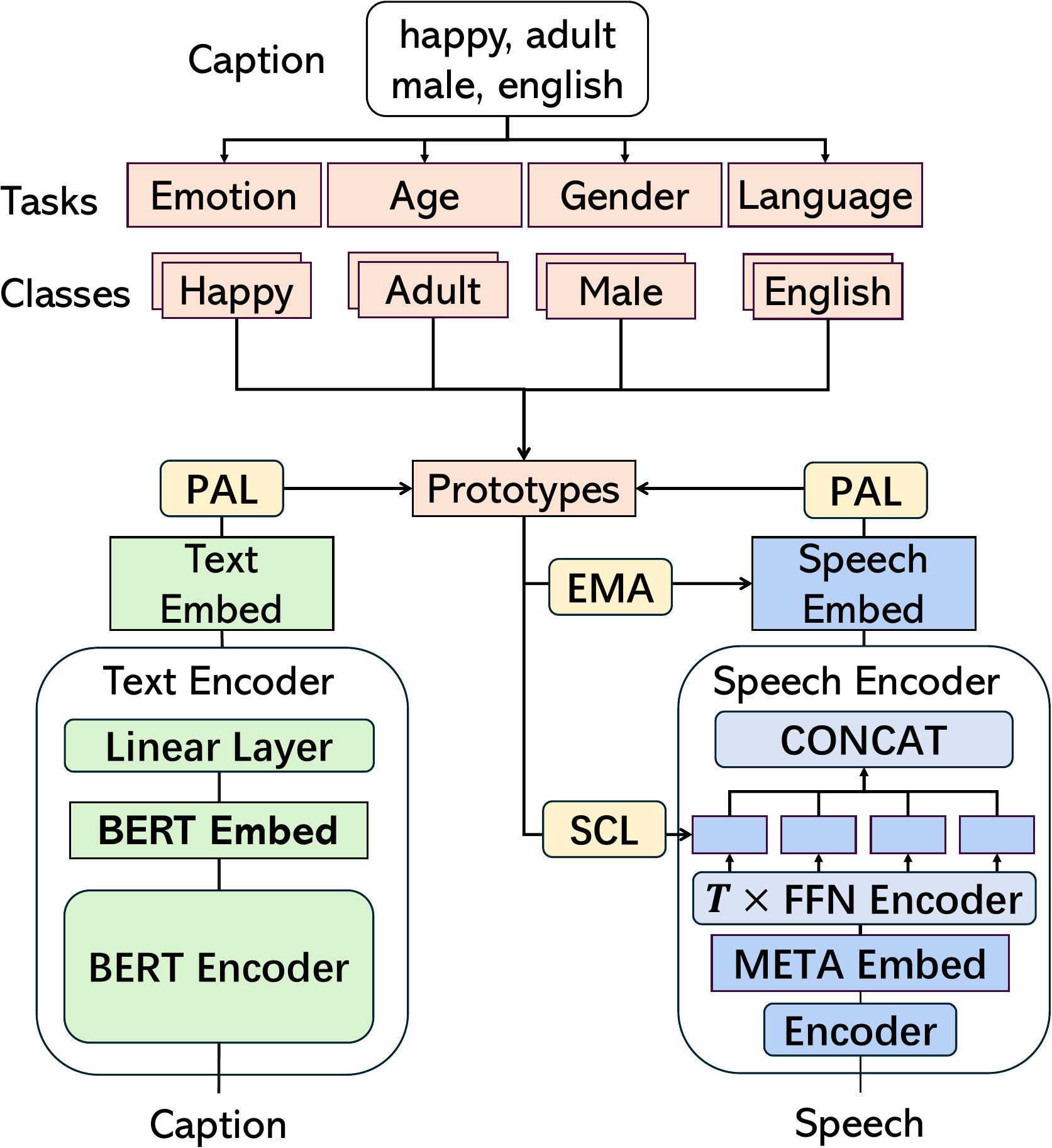}
        \caption{ParaMETA Overview}\label{fig:overall}
    \end{subfigure}
    \hfill
    \begin{subfigure}[b]{0.38\linewidth}
        \includegraphics[width=\linewidth]{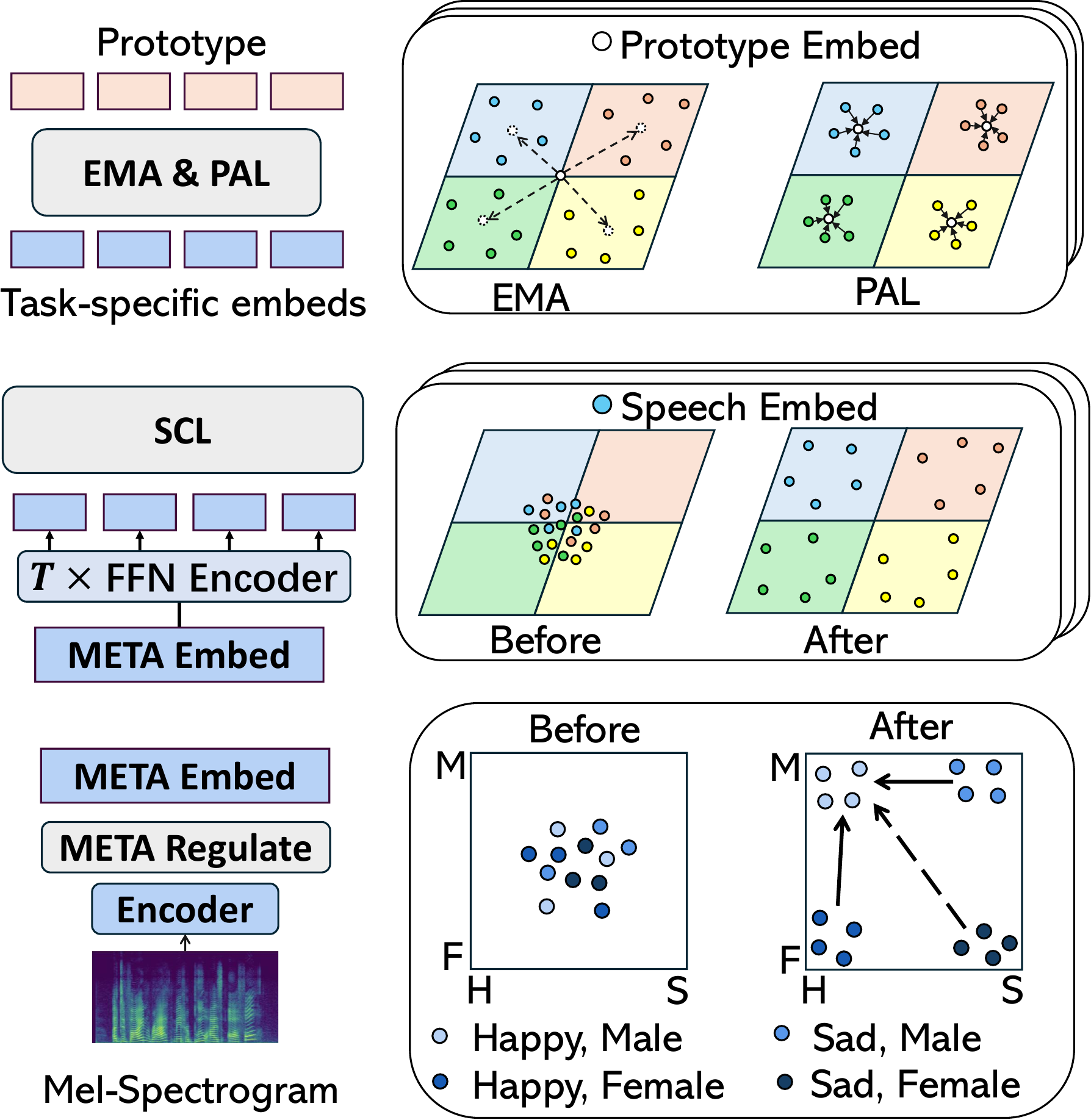}
        \caption{Embedding Spaces}\label{fig:embedding}
    \end{subfigure}
    \hfill
    \begin{subfigure}[b]{0.21\linewidth}
        \includegraphics[width=\linewidth]{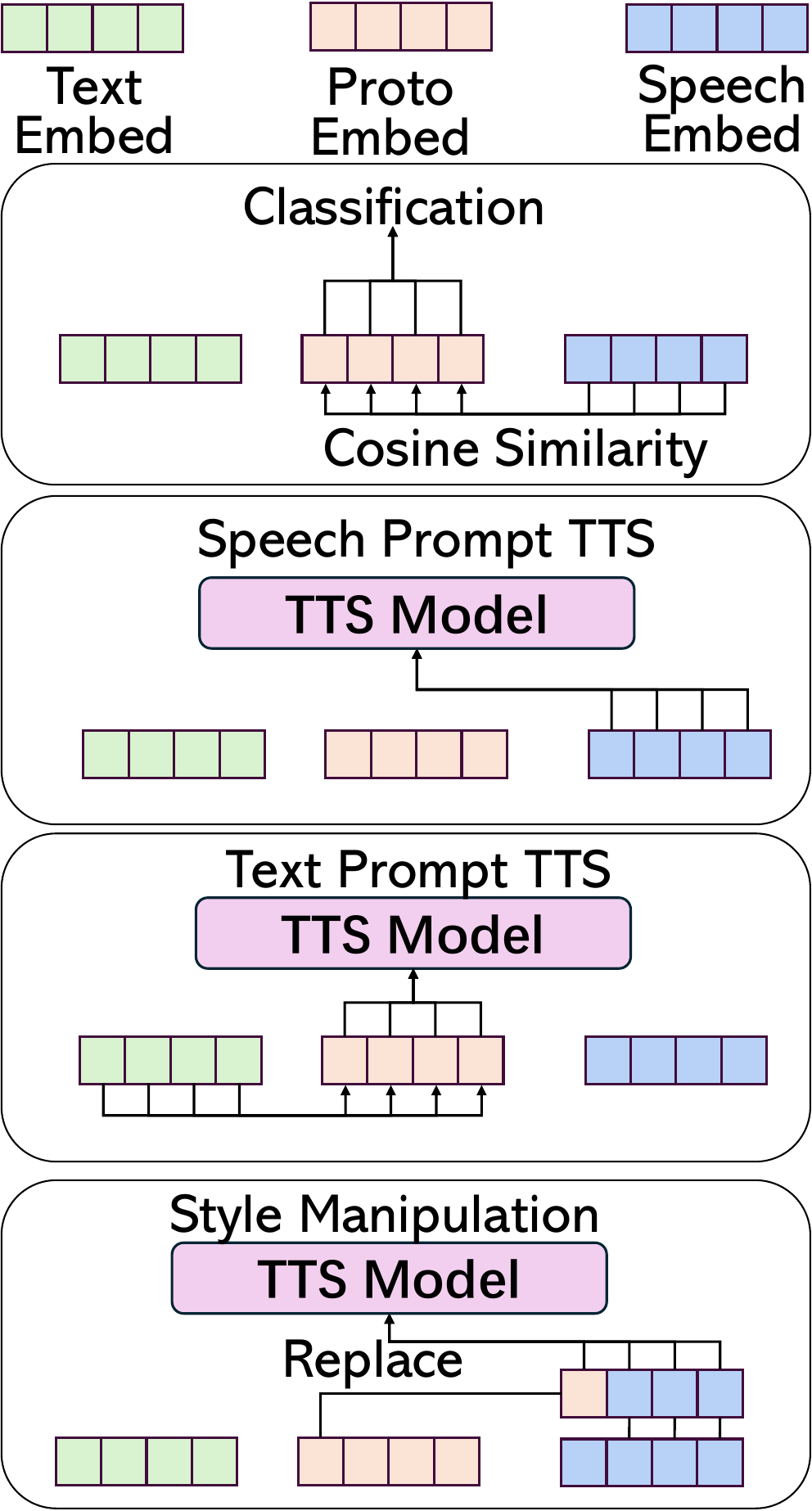}
        \caption{Applications}\label{fig:application}
    \end{subfigure}
    \caption{
      Overview of the \textbf{ParaMETA} framework. Given a speech sample and its style caption, ParaMETA decomposes the caption into $T$ tasks and encodes the speech into a META space with regularization. The META embedding is projected into task-specific subspaces, where contrastive and prototype alignment losses structure the representation. The learned embeddings support downstream applications such as classification, style controllable TTS, and style manipulation via both speech and text prompts.
    }
    \label{fig:parameta_framework}
\end{figure*}

To formalize our terminology, we define a task $T$ as a specific type of speaking styles to be recognized, such as emotion, age, or gender. Each task $t \in T$ is associated with a distinct set of classes~$C_t$, where each class represents a possible value of the corresponding speaking style. For example, happy is a class within the emotion recognition task and does not appear in other tasks, while female is a class within the gender classification task. We assume that each speech is assigned to at most one class per task.
Given a speech sample and its associated label consisting of $T$ tasks, where each task~$t$ has a corresponding class from the set $C_t$, the objective is to learn $T$ task-specific embeddings. 

\subsection{Speech Encoder}
Given a Mel-spectrogram of speech, denoted as $\mathrm{MEL} \in \mathbb{R}^{F \times t}$, where $F$ is the number of frequency bins and $t$ is the number of frames. We define the speech encoding process as:
\begin{equation}
    x = \mathrm{Encoder}(\mathrm{MEL}), x \in \mathbb{R}^{D}
\end{equation}
where $D$ represents the dimensionality of the speech embedding.
ParaMETA is a \textbf{model-agnostic} representation learning framework; we systematically explore a variety of encoder backbone architectures to validate its flexibility and generality. Specifically, we experiment with convolutional neural networks, CNNs, \cite{li2021survey}, long short-term memory networks, LSTMs, \cite{hochreiter1997long}, query-based transformers, Q-Former, \cite{li2023blip}, and standard Transformers~\cite{vaswani2017attention}. 
For the CNN-based encoder, we process $\mathrm{MEL}$ through convolutional layers followed by a global mean pooling over the time dimension to obtain a fixed-length embedding. For the LSTM-based encoder, we use the final hidden state as the sequence representation. For the Q-Former, we introduce a set of learnable latent queries that attend to the spectrogram via cross-attention. For the Transformer-based encoder, we apply self-attention layers over $\mathrm{MEL}$ and use a sum pooling across time steps to produce the final embedding.

\subsection{META Embedding}
Once we collect the speech embedding $X$, we apply the first stage of contrastive learning in the META space.  Traditional contrastive learning frameworks typically require the construction of positive and negative pairs, where positive pairs consist of speeches with exactly the same style labels, and negative pairs consist of samples with any differing labels. This binary formulation treats all non-identical pairs as equally dissimilar, failing to capture the nuanced relationships between partially overlapping speaking styles.

To address this limitation, we adopt a graded similarity framework using a "positive-to-less-positive" contrastive regulation scheme. Instead of treating all negative pairs equally, we define similarity based on the number of shared speaking styles. Speeches that share more speaking styles are considered more similar and are encouraged to be embedded closer in the META space. Intuitively, a speech sample labeled \textit{[female, happy]} should be closer to \textit{[female, sad]} than to \textit{[male, sad]} in the META space, since the former pair shares the same gender, whereas the latter does not.

Specifically, let $X \in \mathbb{R}^{B \times D}$ denote a batch of speech embeddings, where $B$ is the batch size and $D$ is the dimensionality of each embedding. Let $x_i \in \mathbb{R}^D$ represent the $i$-th embedding in the batch. Each speech sample is associated with $T$ task-specific labels corresponding to different speaking styles tasks. For the $i$-th sample, we denote its task-wise labels as $y_i \in \mathbb{Z}^T$, where each $y_i^{(t)}$ indicates the class index for the $t$-th task. The complete set of labels for the batch is denoted by $Y \in \mathbb{Z}^{B \times T}$, where each row $y_i$ corresponds to the annotations for the $i$-th embedding.

For each pair $(i,j)$ of embedding in the batch, we compute a class-wise similarity score $w_{i,j} \in [0,1]$ based on the proportion of shared class. based on the proportion of shared task labels. Specifically, the similarity is defined as the average number of matching labels across all $T$ tasks. We then normalize these similarity scores across the batch (excluding the self-pair) to obtain the final similarity weight $\hat{w}_{i,j}$. 
\begin{equation}
\hat{w}_{i,j} = \frac{w_{i,j}}{\sum_{k \ne i} w_{i,k}}, \quad w_{i,j} = \frac{1}{T} \sum_{t=1}^{T} \mathbb{1}[y_i^{(t)} = y_j^{(t)}],
\end{equation}

We then compute the log-probability that sample $j$ is a positive (i.e., style-similar) neighbor of sample $i$ with a softmax operation over the cosine similarities between $x_i$ and all other embeddings $x_k$, where $k \ne i$ in the batch.
\begin{equation}
   \mathrm{log}p_{i,j} = \mathrm{log}\frac{\mathrm{exp}(\cos \left(x_i,x_j \right))}{\sum_{k \ne i}\mathrm{exp}(\cos \left(x_i,x_k \right))}
\end{equation}
where \( \cos(\cdot, \cdot) \) denotes cosine similarity.
To guide the embedding space such that samples sharing more speaking styles attributes are closer together, we define the META embedding regularization loss $L_{\mathrm{META}}$ as a weighted negative log-likelihood over all sample pairs in the batch:
\begin{equation}
    \mathcal{L}_{\mathrm{META}} = -\frac{1}{B} \sum_{i=1}^{B} \sum_{j\ne i}^{B} \hat{w}_{i,j}\log p_{i,j}
\end{equation}

\subsection{Task-Specific Embedding}
While the shared META space encourages embedding with similar styles to be closer in the space, we further enhance the representation by introducing task-specific structural regularization to reduce the task-wise interference. Specifically, we apply $T$ independent linear projection layers, each mapping the shared META embedding into a dedicated subspace tailored to one task. This design allows each subspace to focus exclusively on features relevant to its corresponding task, thereby reducing interference from unrelated factors and promoting more interpretable and disentangled representations. 

Formally, let \( Z \in \mathbb{R}^{B \times D} \) denote a batch of shared META embeddings. For each  task \( t \in \{1, \dots, T\} \), we define a task-specific linear projection \( f_t: \mathbb{R}^{D} \rightarrow \mathbb{R}^{d} \), which maps the shared embedding into a lower-dimensional task-specific subspace. The resulting task-specific embeddings are denoted by: \( z^{(t)} = f_t(Z) \in \mathbb{R}^{B \times d} \). 
To enhance class-level discriminability within each task, we apply a supervised contrastive loss independently to each task-specific embedding space.  Let \( y_i^{(t)} \in C_t \) denote the class label and  $z_{i}^{(t)}$ represents the embedding of speech sample $i$ for task $t$.  

\[
\mathcal{P}_i^{(t)} = \left\{ j \mid j \ne i,\ y_j^{(t)} = y_i^{(t)} \right\}
\]
All other samples \( j \ne i \) with a different class label are treated as negatives. The supervised contrastive loss for task \( t \) is defined as:

\begin{equation}
\mathcal{L}_{\text{SCL}}^{(t)}
= -\frac{1}{B}\!\sum_{i=1}^{B}\!
\frac{1}{|\mathcal{P}_i^{(t)}|}\!
\sum_{j\in\mathcal{P}_i^{(t)}}
\log\frac{e^{\cos(z_i,z_j)}}{\sum_{k\neq i} e^{\cos(z_k,z_j)}}
\end{equation}

This objective encourages embeddings of the same class to cluster closely together within each task-specific subspace, while pushing apart those from different classes. The total supervised contrastive loss across all tasks is given by:
\begin{equation}
\mathcal{L}_{\mathrm{SCL}} = \sum_{t=1}^{T} \mathcal{L}_{\mathrm{SCL}}^{(t)}
\end{equation}

This formulation promotes intra-class compactness and inter-class separability within each task-specific embedding space, while maintaining independence across all $T$ tasks.

\subsection{Prototype Learning}
While contrastive learning helps structure the embedding space, it does not directly provide class representations for downstream tasks such as classification or control. To bridge this gap, we introduce prototype embeddings that act as class anchors within each task-specific subspace. 

For each task \( t \), we define a prototype matrix \( P^{(t)} \in \mathbb{R}^{C_t \times d} \), where \( C_t \) is the number of classes for task \( t \), and \( p_c^{(t)} \in \mathbb{R}^{d} \) represents the prototype for class \( c \). For example, \( p_{\text{adult}}^{(\text{age})} \) denotes the prototype corresponding to the ``adult'' class in the age classification task. 
These prototypes are randomly initialized, but during training, an \textbf{E}xponential \textbf{M}oving \textbf{A}verage (EMA) update is applied to gradually move each prototype toward the centroid of the corresponding class in the current batch. This dynamic update strategy ensures that the prototypes evolve smoothly to reflect the underlying distribution of task-specific embeddings. 

Formally, given a batch of task-specific embeddings \( Z^{(t)} \in \mathbb{R}^{B \times d} \) and their corresponding class labels \( y_i^{(t)} \), we compute the \textit{centroid} of each class \( c \in \{1, \dots, C_t\} \) within the batch as:
\[
z_c^{(t)} = \frac{1}{|\mathcal{I}_c^{(t)}|} \sum_{i \in \mathcal{I}_c^{(t)}} z_i^{(t)}, \quad \text{where } \mathcal{I}_c^{(t)} = \left\{ i \mid y_i^{(t)} = c \right\}
\]
Here, \( z_c^{(t)} \in \mathbb{R}^d \) represents the average embedding of samples assigned to class \( c \) in task \( t \) within the current batch.

The prototype for class \( c \) in task \( t \), denoted \( p_c^{(t)} \in \mathbb{R}^d \), is then updated using an \textit{exponential moving average (EMA)} to gradually incorporate new information while preserving stability~\cite{he2020momentum}:
\[
p_c^{(t)} \leftarrow m \cdot p_c^{(t)} + (1 - m) \cdot z_c^{(t)}
\]
where \( m \in [0, 1) \) is the momentum coefficient that controls the update rate. In our experiments, we set \( m = 0.99 \) to ensure smooth and stable prototype evolution over time.

While prototypes are designed to reflect the distributional centroids of speech embeddings, we also encourage the embeddings themselves to align with their corresponding prototypes. To achieve this, we introduce a \textit{prototype alignment loss} \( \mathcal{L}_{\mathrm{PAL}} \), which pulls each sample's embedding toward its class prototype within the appropriate task-specific subspace.  For a given task \( t \), let \( z_i^{(t)} \in \mathbb{R}^d \) be the task-specific embedding of sample \( i \), and let \( p_{c}^{(t)} \in \mathbb{R}^d \) be the prototype corresponding to its class label \( c \) and overall prototype alignment loss~$\mathcal{L}_{\mathrm{PAL}}$ across all tasks is:

\begin{equation}
\mathcal{L}_{\mathrm{PAL}} = \sum_{t=1}^{T} \frac{1}{B} \sum_{i=1}^{B} \left( 1 - \cos\left(z_i^{(t)},\ p_{c}^{(t)}\right) \right)
\end{equation}

This objective encourages each embedding to move closer to its semantic center within the task-specific space, thereby reinforcing class-level consistency and enhancing the discriminative structure of the learned representations. Fig.\ref{fig:embedding} provides an intuition of how different objectives shape the structure of the embedding space.

\subsection{Text and Speech Alignment}
To enable a unified framework that supports both speech-based and text-based style control, we incorporate style captions that describes the intended speaking style. Each caption is first encoded using a pretrained text encoder~\cite{song2020mpnet}, and the resulting embedding is then projected into the corresponding task-specific subspace.

To ensure semantic consistency across modalities, we apply the same prototype alignment loss to the projected text embedding. Specifically, we minimize the distance between the projected embedding and the prototype of the target style class referred to by the caption. This alignment allows the TTS model to support both speech and text prompts for style control within a unified framework.

The final training objective combines the META space regularisation, task-specific supervised contrastive loss, and prototype alignment losses for both speech and text modalities:
\begin{equation}
\mathcal{L} = \mathcal{L}_{\mathrm{META}} + \mathcal{L}_{\mathrm{SCL}} + \mathcal{L}_{\mathrm{PAL}}^{(\mathrm{Speech})} +  \mathcal{L}_{\mathrm{PAL}}^{(\mathrm{Text})}
\end{equation}

\begin{table*}[htbp]
\centering
\resizebox{\textwidth}{!}{
\begin{tabular}{ll|ccc|ccc|ccc|ccc}
\toprule
\multirow{2}{*}{} & \multirow{2}{*}{} &
\multicolumn{3}{c|}{Emotion} &
\multicolumn{3}{c|}{Gender} &
\multicolumn{3}{c|}{Age} &
\multicolumn{3}{c}{Language}  \\
\cmidrule(lr){3-5} \cmidrule(lr){6-8} \cmidrule(lr){9-11} \cmidrule(lr){12-14}
& & B. Acc & Macro F1 & W. F1 & B. Acc & Macro F1 & W. F1 & B. Acc & Macro F1 & W. F1 & B. Acc & Macro F1 & W. F1 \\
\midrule
\multicolumn{2}{l}{Baselines} \\
\midrule
CLAP & General  & 14.3   &  0.036  & 0.036 & 50.0   &  0.333  & 0.333 & 25.0   &  0.100  & 0.100 & 50.0   &  0.333  & 0.333 \\
CLAP & Speech \& Music & 22.1   &  0.149  & 0.170 & 67.1   &  0.457  & 0.685 & 11.9   &  0.070  & 0.105 & 18.9   &  0.143  & 0.215 \\
ParaCLAP &  & 9.2   &  0.049  & 0.056 & 9.7   &  0.098  & 0.147 & 10.8   &  0.085  & 0.123 & 20.0   &  0.205  & 0.307 \\
\midrule
Backbone & Objective \\
\midrule
CNN &  Cross Entropy & \textbf{15.1}   &  \textbf{0.082}  & \textbf{0.093} & 63.2 & 0.591  & 0.591  & 20.9 & 0.072  & 0.090  & 61.9 & 0.618  & 0.618  \\
CNN &  CLAP & 13.4 & 0.050  & 0.057  & 59.6 & 0.382  & 0.574  & 16.4   &  \textbf{0.112}  & \textbf{0.168} & 48.7 & 0.356  & 0.533  \\
CNN & ParaMETA (w/o r)  & 14.8 & 0.049  & 0.056  & 61.8 & 0.561  & 0.561  & 20.6 & 0.076  & 0.095  & \textbf{71.8}   &  \textbf{0.718}  & \textbf{0.718} \\
CNN &  ParaMETA (w r)  & 15.0 & 0.056  & 0.064  & \textbf{64.2}   &  \textbf{0.602}  & \textbf{0.602} & \textbf{21.7} &  0.088   &  0.109  & 68.5 & 0.685  & 0.685  \\     
\midrule
LSTM &  Cross Entropy  & \textbf{61.3}   &  \textbf{0.613}  & \textbf{0.613} & 76.8 & 0.761  & 0.761  & 23.5 & 0.138  & 0.172  & 92.3 & 0.923  & 0.923  \\ 
LSTM &  CLAP & 57.1 & 0.495  & 0.565  & 66.6 & 0.470  & 0.705  & 25.1  &  0.197   & 0.296 & 49.7 & 0.426  & 0.639  \\ 
LSTM & ParaMETA (w/o r)  & 51.6 & 0.488  & 0.513  & \textbf{77.2}   &  \textbf{0.764}  & \textbf{0.764} & \textbf{36.3}  & \textbf{0.237} &   \textbf{0.296}  & 92.5 & 0.925  & 0.925  \\
LSTM &  ParaMETA (w r) & 49.0 & 0.489  & 0.489  & 76.2 & 0.754  & 0.754  & 32.8 & 0.233  & 0.291  & \textbf{92.8}   &  \textbf{0.928}  & \textbf{0.928} \\
\midrule
QFormer &  Cross Entropy & 46.0 & 0.451  & 0.451  & 72.1 &  0.710   &  0.710  & 21.4 & 0.115  & 0.144  & 81.1 & 0.810  & 0.810   \\ 
QFormer &  CLAP & \textbf{53.2}   &  \textbf{0.520}  & \textbf{0.520} & 35.4 & 0.328  & 0.491  & 18.2 & 0.168  & 0.252  & 48.6 & 0.397  & 0.595  \\
QFormer &  ParaMETA (w/o r) & 47.2 & 0.470  & 0.470  & \textbf{72.1}   &  \textbf{0.712}  & \textbf{0.712} & \textbf{38.2}   &  \textbf{0.264}  & \textbf{0.330} & \textbf{81.8}   &  \textbf{0.818}  & \textbf{0.818} \\
QFormer &  ParaMETA (w r) & 44.1 & 0.423  & 0.423  & 71.7 & 0.711  & 0.711  & 35.8 & 0.262  & 0.328  & 78.9 & 0.789  & 0.789  \\
\midrule
Transformer & Cross Entropy & 35.0 & 0.285  & 0.308  & 76.8 & 0.760  & 0.760  & 20.6 & 0.092  & 0.114  & 89.5 & 0.895  & 0.895  \\ 
Transformer &  CLAP & \textbf{55.2}   &  \textbf{0.528}  & \textbf{0.541} & 39.4 & 0.353  & 0.529  & 25.3   &  \textbf{0.210}  & \textbf{0.316} & 56.6 & 0.456  & 0.684  \\
Transformer & ParaMETA (w/o r) & 44.2 & 0.404  & 0.427  & 77.9 & 0.773  & 0.773  & 26.1 & 0.171  & 0.214  & 90.7 & 0.907  & 0.907  \\
Transformer & ParaMETA (w r) & 50.1 & 0.501  & 0.501  & \textbf{78.4}   &  \textbf{0.778}  & \textbf{0.778} & \textbf{29.7} &  0.194   &  0.243  & \textbf{91.1}   &  \textbf{0.910}  & \textbf{0.910} \\
\bottomrule
\end{tabular}
}
\caption{Subject-Independent Classification Performance Evaluation. The evaluation is performed with multiple runs.}
\label{tab:class_performance_metrics}
\end{table*}

\subsection{Downstream Application}
The proposed \textbf{ParaMETA} framework enables a variety of downstream applications by leveraging its disentangled, task-specific paralinguistic embeddings and unified prototype-aligned structure.  Fig.\ref{fig:application} illustrates how different tasks are handled within the unified ParaMETA framework.
\paragraph{Speaking Styles Classification.}
Given a speech input, we extract its shared META embedding and obtain task-specific representations through projection heads. Each task-specific embedding is then matched to the most similar class prototype using cosine similarity. The predicted class is the one with the highest similarity score in the corresponding task subspace.
\paragraph{Style Controllable TTS.}
ParaMETA supports both speech-based and text-based style control for TTS generation. In the speech-to-speech setting, task-specific embeddings are concatenated and used to condition the TTS model, enabling the generated speech to mimic the speaking styles of a reference speeches. In the text-based setting, a caption describing the desired style (e.g., "happy adult female") is encoded and projected into the prototype space. The resulting embedding is aligned with the corresponding style prototypes, allowing the model to generate speech conditioned solely on textual descriptions.

\paragraph{Speaking Styles Manipulation.}
Thanks to its disentangled embedding structure, the ParaMETA framework not only enables TTS models to replicate the speaking styles from a reference speech but also supports fine-grained control over a specific type of speaking style. For example, given a speech with the style \textit{[male, happy, adult]}, ParaMETA outputs a disengaged embedding for each type of speaking style, gender, emotion, and age. By replacing the emotion embedding (e.g., from happy to the prototype of sad), we can modify the emotional style while keeping other styles unchanged.

\section{Experiments}
\subsection{Dataset and Experimental Setup }
We construct a multilingual and multi-style speech dataset by combining multiple open-source corpora to ensure broad coverage across paralinguistic tasks and class labels. Specifically, we combine: \textbf{Baker} \cite{BakerDataset2020}, \textbf{LJSpeech} \cite{ljspeech17}, \textbf{ESD} \cite{zhou2021seen}, \textbf{CREMA-D} \cite{cao2014crema} and characters voices from  publicly available \textbf{Genshin Impact} dataset.  The combined dataset covers 16 speaking styles: Emotion (happy, angry, sad, neutral, surprise, disgust, fear), Age (child, teenager, young adult, adult, senior), Gender (male, female), and Language (English, Chinese) and approximately 93k speech samples. All speeches resampled to a standardized sampling rate of 22.05 kHz.
Training is conducted on NVIDIA TITAN RTX GPUs using a batch size of 32 for up to 40k steps. Model is optimized using the AdamW optimizer~\cite{loshchilov2017decoupled} with a learning rate of 2e-4.

\subsection{Speaking Styles Classification}
We evaluate state-of-the-art open-source baselines for the classification task, including \textbf{CLAP}\cite{elizalde2023clap} and \textbf{ParaCLAP}\cite{jing2024paraclap}. For CLAP, we compare both the general-purpose model and a version pretrained on music and speech. ParaCLAP, specifically designed for paralinguistic tasks, is also evaluated using its pretrained model. To generate text embeddings, we input the corresponding class labels of each speech sample in our dataset into the pretrained text encoder. Classification is then performed by computing the similarity between the speech embedding and all candidate text embeddings, assigning the class with the highest similarity score.
We compare ParaMETA's classification performance against other commonly used representation learning frameworks in learning speaking styles. Specifically, based on the same speech encoding backbone, we implement a multi-task classifier trained with cross-entropy loss as a conventional baseline.   We also train a model that learns to align text and speech embeddings through a contrastive loss, following the objective used in CLAP.  To isolate the effect of META embedding regularization, we additionally conduct an ablation study comparing models trained with and without the $\mathcal{L}_{\mathrm{META}}$ loss. All evaluations are conducted under a strict subject-independent setting, where speakers in the training and testing sets are disjoint. Each evaluation is repeated across five runs with different sampled test sets to ensure a fair comparison.
\begin{figure}
    \centering
    \begin{subfigure}[b]{0.45\linewidth}
        \centering
        \includegraphics[width=\linewidth]{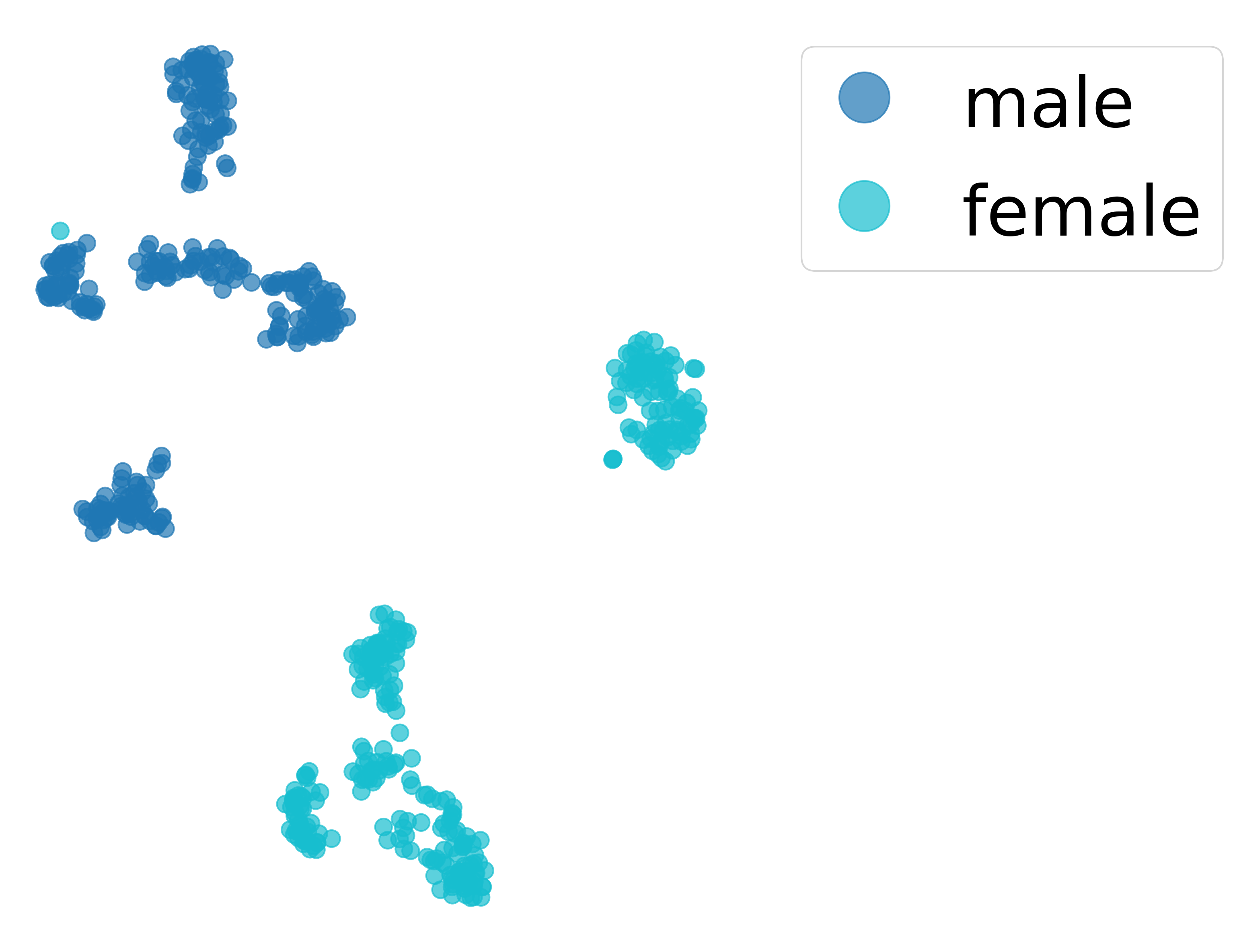}
        \caption{META-Gender}
        \label{fig:gender_mata}
    \end{subfigure}
    \hfill
    \begin{subfigure}[b]{0.45\linewidth}
        \centering
        \includegraphics[width=\linewidth]{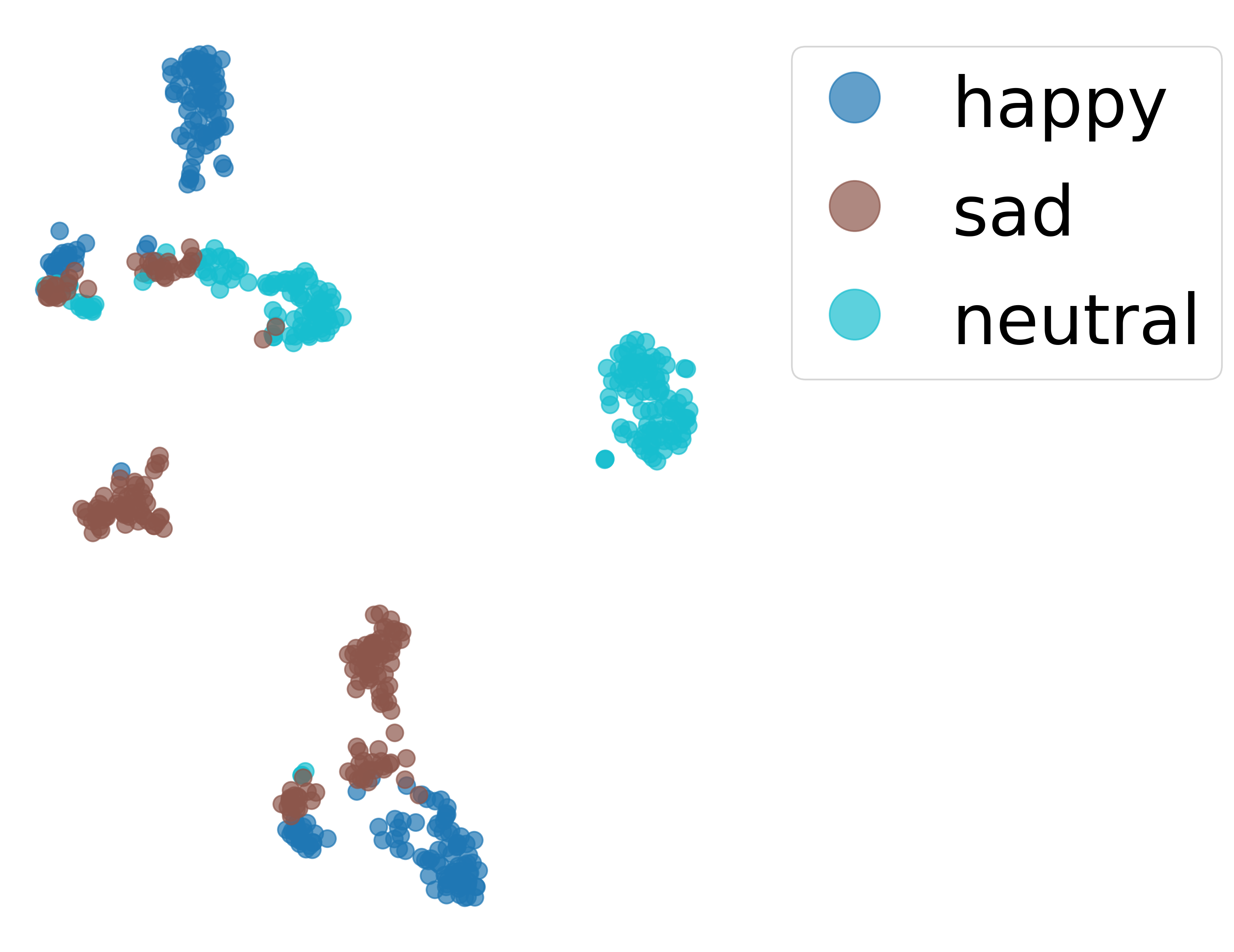}
        \caption{META-Emotion}
        \label{fig:emotion_mata}
    \end{subfigure}
    \hfill
    \begin{subfigure}[b]{0.45\linewidth}
        \centering
        \includegraphics[width=\linewidth]{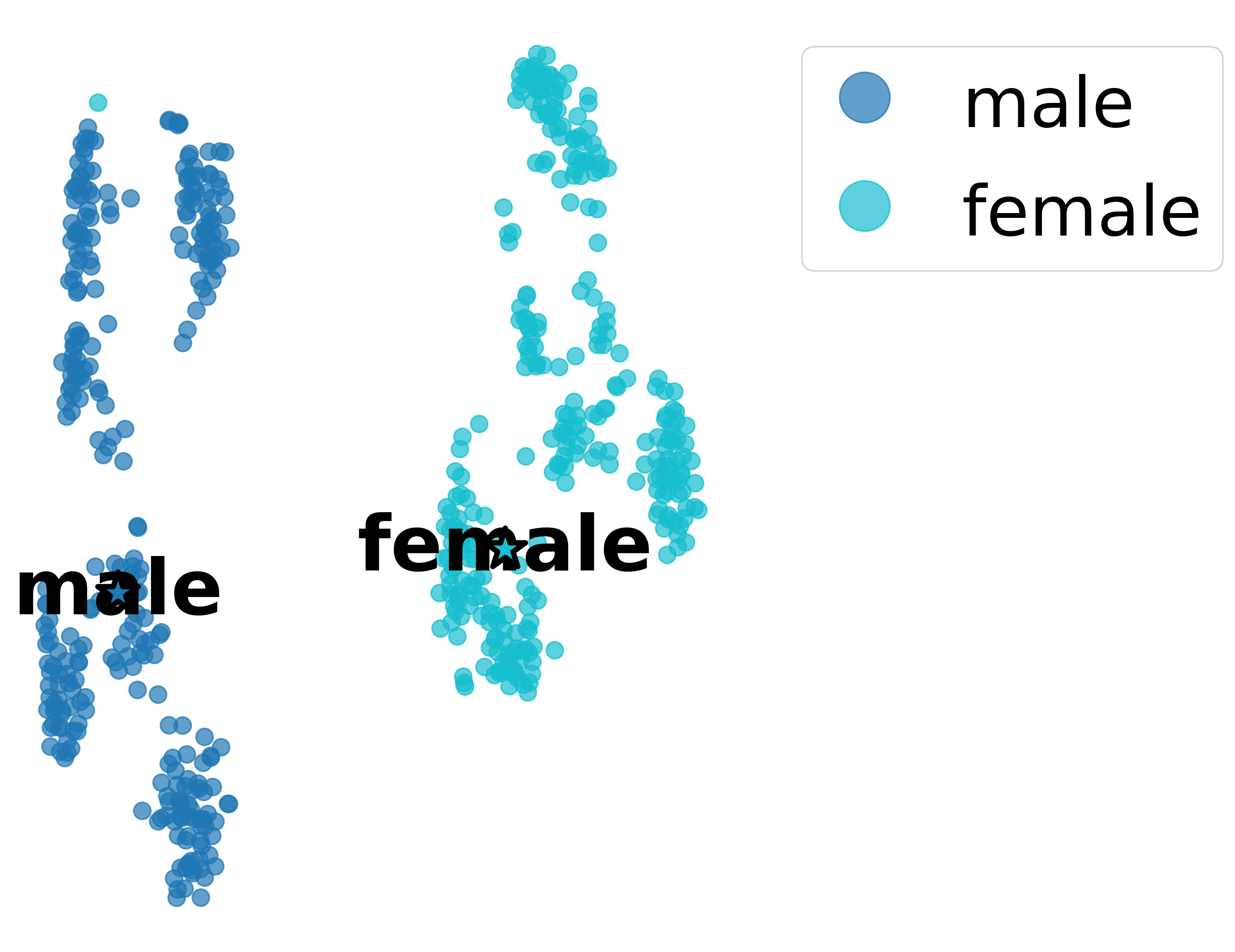}
        \caption{Gender-Specific}
        \label{fig:gender_specific}
    \end{subfigure}
    \hfill
    \begin{subfigure}[b]{0.45\linewidth}
        \centering
        \includegraphics[width=\linewidth]{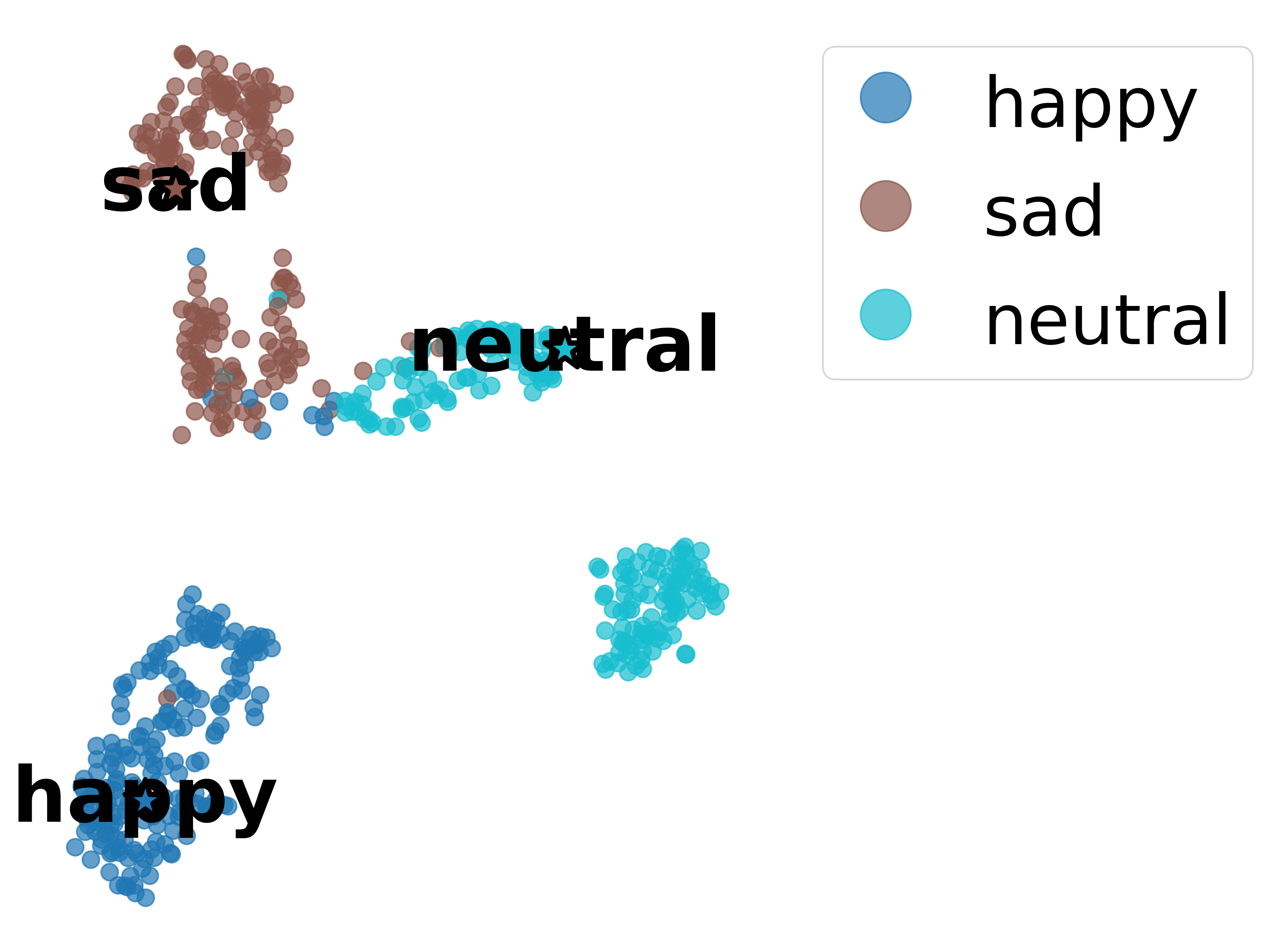}
        \caption{Emotion-Specific}
        \label{fig:emotion_specific}
    \end{subfigure}
    \hfill
    \caption{Embedding Visualization}
    \label{fig:embedding_visualize}
\end{figure}

Table \ref{tab:class_performance_metrics} reports the subject-independent classification results across four speaking-style recognition tasks. Models pretrained on large-scale multimodal datasets (CLAP and ParaCLAP) show consistently poor performance across all tasks. For instance, the general CLAP model achieves only 14.3\% balanced accuracy on emotion, 50\% on gender, 25\% on age, and 50\% on language. ParaCLAP performs even worse, with emotion at 9.2\%, gender at 9.7\%, age at 10.8\%, and language at 20\%.

Training encoders directly on our dataset significantly improves performance relative to pretrained models. However, these methods still exhibit \textbf{negative transfer}. Cross-entropy training, for example, performs well on LSTM-based emotion and language classification but shows remarkable drops in age accuracy across backbones (e.g., a 10–18\% reduction compared to ParaMETA). 
Models trained with CLAP-style objectives suffer the most from negative transfer. Their performance are generally strong in emotion recognition but collapses greatly on gender, age, and language recognition. Cross-entropy mitigates this issue to some extent and produces competitive results in several settings, but it remains still exhibits nontrivial negative transfer and noticeably weaker than ParaMETA on fine-grained attributes such as age. 

Across all backbones, ParaMETA delivers the most stable and uniformly strong performance. Out of 16 backbone–task combinations, ParaMETA achieves the best result in 12 cases, indicating that its embedding space is more discriminative, robust, and less susceptible to inter-task interference. These results highlight ParaMETA's strength in learning useful, task-relevant, and partially disentangled speaking-style representations directly from speech. The performance differences can be explained by how each training objective structures the embedding space. CLAP-style objectives force speech embeddings to align with text embeddings, treating the entire speech as a single semantic unit. This produces highly entangled representations, where paralinguistic (emotion, gender, age, language) are compressed into a shared latent space dominated by whichever features correlate most with text. 
Cross-entropy training improves disentanglement because it learns each paralinguistic label as an independent prediction task. This reduces entanglement relative to CLAP, but the model must implicitly discover and separate all style factors without explicit constraints. Since there is no mechanism to control or structure the embedding space, the resulting representations can still be unstable, sensitive to backbone choice, and prone to inter-task interference. ParaMETA addresses these limitations by introducing task-specific prototype embeddings and META-space regularization, which explicitly guide the encoder to learn separable latent subspaces for different tasks. Because each task anchors its own prototype and remains independent from others, the model avoids cross-task suppression and achieves more stable, robust, and disentangled representations. Fig.\ref{fig:embedding_visualize} shows t-SNE visualizations of the learned speaking styles embeddings.  To illustrate the structure of the embedding space, we show examples from two representative tasks: gender (male, female) and emotion (happy, sad, neutral). As shown in Figures~\ref{fig:gender_mata} and~\ref{fig:emotion_mata}, the META space exhibits distinct clusters, but the structure is primarily shaped by gender. For instance, a happy male embedding appears closer to a sad male than to a happy female.  It suggests that gender overrides emotion in organizing the space. This overlap makes it difficult to learn emotion-specific prototypes due to interference from gender task. In contrast, the task-specific embeddings in Figures~\ref{fig:gender_specific} and~\ref{fig:emotion_specific} show clearer separation by emotion, with reduced influence from gender. 

While ParaMETA consistently outperforms other objectives, the role of META-space regularization merits brief discussion. Although not universally beneficial, it improves performance in 8 of 16 backbone–task settings. For most backbones, the difference between regularized and non-regularized variants is small, typically within 3\%. The Transformer backbone, however, shows remarkable larger gains: META regularization boosts emotion accuracy by 6\% and age accuracy by 3.6\%, both of which are more challenging multi-class tasks. This suggests that META regularization is particularly effective when finer style separation is required. Although the evidence is not yet conclusive, it indicates clear potential for improving representation quality in difficult settings, and further analysis will be pursued in future work.

\begin{table}[t]
\centering
\begin{tabular}{l|ccc}
\toprule
Prompt Type & N-MOS & E-MOS \\
\midrule
Text Only & 2.02  $\pm$ 0.69 & 2.33 $\pm$ 0.97 \\ 
Speech Only & 2.89  $\pm$ 0.82 & 3.19 $\pm$ 0.88 \\ 
\midrule
ParaMETA Text & 3.06  $\pm$ 0.71 & 2.91 $\pm$ 0.87 \\ 
ParaMETA Speech &  \textbf{3.41  $\pm$ 0.86} & \textbf{3.41 $\pm$ 1.10} \\ 
\bottomrule
\end{tabular}
\caption{Speech Generative Model Evaluation}
\label{tab:tts_performance}
\end{table}
\subsection{Style-Controllable TTS Generation}
For generative evaluation, we investigate how different types of speaking-style embeddings influence the perceptual quality of synthesized speech. We train ParaStyleTTS~\cite{10.1145/3746252.3761311}, a multilingual and style-controllable TTS model conditioned on a style embedding $Style \in \mathbb{R}^{D}$. Three variants of ParaStyleTTS are developed: one using text-based style embeddings, one using speech embeddings learned jointly through a Transformer encoder during TTS training, and one using speech embeddings produced by a Transformer encoder pretrained with the ParaMETA. To assess the perceptual impact of each embedding type, we conduct a subjective listening study in which multiple speech samples are generated and evaluated by five multilingual speakers, who rate the perceived naturalness (N-MOS) and expressiveness (E-MOS) of the generated speech.

Table~\ref{tab:tts_performance} presents the evaluation results. The scores that show statistically significant improvements based on a two-tailed paired t-test ($p < 0.05$) are highlighted in bold.
TTS models conditioned on ParaMETA embeddings consistently achieve higher perceptual quality across both text- and speech-based prompting. In general, speech-based prompting yields better results than text-based prompting, with N-MOS and E-MOS scores of 2.89 and 3.19, respectively, compared to 2.00 and 2.33 for text prompts. This performance gap exists because speech contains rich information such as pitch, speed, and tone that naturally convey how a speaking styles sounds. In contrast, text-based prompts like "happy male" are more ambiguous. The same description of speaking styles can be spoken in many different ways depending on the speaker and context, making it harder for the model to generate consistent and expressive speech from text alone.
ParaMETA embeddings offer notable improvements in both prompting methods. For text-based prompts, ParaMETA yields an improvement of 1.0 in naturalness and 0.6 in expressiveness. For speech-based prompts, they yield an improvements of 0.5 in naturalness and 0.2 in expressiveness.  These gains are attributed to the speaking styles disentangled nature of ParaMETA embeddings. , which capture clean, style-relevant representations. In contrast, raw speech embeddings may contain irrelevant information such as background noise, silences, or unintended details. Incorporating such details into the TTS model can hinder performance, while ParaMETA embeddings focus solely on the intended speaking style, resulting in more natural and expressive generation

\subsection{Speaking Styles Manipulation}
\begin{table}[tbp]
\centering
\begin{tabular}{l|cccc}
\toprule
& Orig. Sim. & Manip. Sim. & Accuracy \\
\midrule
Language& 0.4812 & 0.4850 & 55.00  \\
Age & 0.4707 & 0.5486 & 70.00 \\
Gender & 0.4707 & 0.9888 & 100.00 \\
Emotion & 0.4687 & 0.8367 & 90.00 \\
\bottomrule
\end{tabular}
\caption{speaking styles Manipulation}
\label{tab:manipuilation}
\end{table}

We test ParaMETA’s ability to manipulate speaking styles to assess its controllability in TTS model. Given a speech sample, we change one type of speaking styles (e.g., gender or emotion) while keeping others unchanged. We measure the similarity between the manipulated embedding and the target style prototype (higher values indicate better alignment), the similarity between the prototype and the original style, and the classification accuracy by checking whether the manipulated embedding is identified as the target style instead of the original.
As shown in Table~\ref{tab:manipuilation}, the framework enables strong and targeted control. Gender manipulation achieves perfect accuracy (100\%) with a large similarity shift (0.47 to 0.99), indicating high precision. Emotion and age also show strong and moderate effectiveness (90\% and 70\%), respectively. Language manipulation, however, results in minimal change (similarity shift from  0.481 to 0.485, accuracy 55\%), suggesting limited controllability. This is likely because language is primarily expressed through phonetic and lexical content, which is tightly bound to the input text. As such, changing language identity without modifying the text has little effect

\subsection{Computational Resource Comparision}
\begin{table}[tbp]
\centering
\begin{tabular}{l|ccc}
\toprule
Method & RTF & \# Parameters & CUDA Memory\\
\midrule
CLAP & 0.091 & 198.48 M& 1966 MB \\
ParaCLAP & 0.008 & 276.33 M  & 1345 MB \\
\midrule
CNN & 0.002 & 1.94 M & 456 MB\\
LSTM & 0.003 & 3.77 M & 433 MB \\
QFormer & 0.006 & 2.38 M & 439 MB\\
Transformer & 0.005 & 1.86 M & 429 MB \\
\bottomrule
\end{tabular}
\caption{Computational Resource Comparison}
\label{tab:resource_comparision}
\end{table}
Last, we evaluate the computational efficiency of \textbf{ParaMETA} in terms of model size, memory usage, and inference speed. Table~\ref{tab:resource_comparision} compares ParaMETA against baseline models across different backbones. ParaMETA achieves substantial efficiency gains. The LSTM-based variant runs 30$\times$ faster than CLAP (RTF 0.003 vs. 0.091) and is 2.7$\times$ faster than ParaCLAP. It also requires only 3.77 million parameters, roughly 1.3\% of ParaCLAP and 1.9\% of CLAP, indicating a significantly smaller model footprint. GPU memory consumption remains consistently low across ParaMETA backbones, averaging around 440~MB, compared to 1345~MB for ParaCLAP and 1966~MB for CLAP, resulting in a 70\% reduction in CUDA memory usage. These results demonstrate that ParaMETA not only improves style controllability and accuracy, but also offers lightweight, fast, and resource-efficient performance, making it favorable for deployment on resource-constrained or real-time applications.

\section{Conclusion}
In conclusion, we present ParaMETA, a unified representation learning framework designed to disentangle paralinguistic speaking styles. ParaMETA projects speech into structured, task-specific subspaces, enabling interpretable, modular, and interference-reduced style representations learning. By projecting each type of style into distinct subspaces, ParaMETA effectively reduces inter-task interference and supports multi-task paralinguistic recognition with minimal negative transfer.
Beyond recognition, ParaMETA also improves the controllability of TTS. It supports both speech- and text-based prompting and enables precise, expressive, and natural speaking style manipulation.

\bibliography{aaai2026}

\end{document}